\documentclass[final]{cvpr}

\usepackage{times}
\usepackage{epsfig}
\usepackage{graphicx}
\usepackage{amsmath}
\usepackage{amssymb}
\usepackage{booktabs}
\usepackage{multirow}
\usepackage{url}

\usepackage[pagebackref=true,breaklinks=true,colorlinks,bookmarks=false]{hyperref}

\def\cvprPaperID{****} 
\def\confYear{CVPR 2023}

\begin{document}

\title{MIPI 2023 Challenge on RGBW Remosaic: Methods and Results}

\author{
Qianhui Sun \and Qingyu Yang \and Chongyi Li \and Shangchen Zhou \and Ruicheng Feng \and Yuekun Dai \and Wenxiu Sun \and Qingpeng Zhu \and Chen Change Loy \and Jinwei Gu \and
Yuqing Liu \and Hongyuan Yu \and Weichen Yu \and Zhen Dong \and Binnan Han \and Qi Jia \and Xuanwu Yin \and Kunlong Zuo \and
Yaqi Wu \and Zhihao Fan \and Fanqing Meng \and Xun Wu \and Jiawei Zhang \and Feng Zhang \and
Mingyan Han \and Jinting Luo \and Qi Wu \and Ting Jiang \and Chengzhi Jiang \and Wenjie Lin \and Xinpeng Li \and Lei Yu \and Haoqiang Fan \and Shuaicheng Liu
}

\maketitle

\begin{abstract}
Developing and integrating advanced image sensors with novel algorithms in camera systems are prevalent with the increasing demand for computational photography and imaging on mobile platforms.
However, the lack of high-quality data for research and the rare opportunity for an in-depth exchange of views from industry and academia constrain the development of mobile intelligent photography and imaging (MIPI).
With the success of the \href{https://mipi-challenge.org/MIPI2022/}{1st MIPI Workshop@ECCV 2022}, we introduce the second MIPI challenge, including four tracks focusing on novel image sensors and imaging algorithms.
This paper summarizes and reviews the RGBW Joint Remosaic and Denoise track on MIPI 2023.
In total, 81 participants were successfully registered, and 4 teams submitted results in the final testing phase.
The final results are evaluated using objective metrics, including PSNR, SSIM, LPIPS, and KLD. 
A detailed description of the top three models developed in this challenge is provided in this paper. 
More details of this challenge and the link to the dataset can be found at \href{https://mipi-challenge.org/MIPI2023/}{https://mipi-challenge.org/MIPI2023/}.
\end{abstract}

{\let\thefootnote\relax\footnotetext{%
\tiny  Qianhui Sun$^{1}$ (\href{sunqianhui@sensebrain.site}{sunqianhui@sensebrain.site}), Qingyu Yang$^{1}$ (\href{yangqingyu@sensebrain.site}{yangqingyu@sensebrain.site}), Chongyi Li$^{4}$, Shangchen Zhou$^{4}$, Ruicheng Feng$^{4}$, Wenxiu Sun$^{2,3}$, Qingpeng Zhu$^{2}$, Chen Change Loy$^{4}$, Jinwei Gu$^{1,3}$ are the MIPI 2023 challenge organizers
($^{1}$SenseBrain, $^{2}$SenseTime Research and Tetras.AI, $^{3}$Shanghai AI Laboratory, $^{4}$Nanyang Technological University). The other authors participated in the challenge. Please refer to Appendix~\ref{appendix:teams} for details.
\\
MIPI 2023 challenge website: \href{https://mipi-challenge.org/MIPI2023/}{https://mipi-challenge.org/MIPI2023/}
}
}


\section{Introduction}
\begin{figure}[!ht]
\centering
\includegraphics[width=0.5\textwidth]{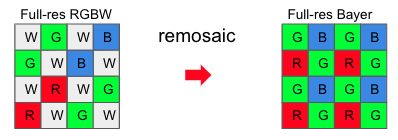}
\caption{The RGBW remosaic task.}
\label{fig:rgbw_cfa}
\setlength{\belowcaptionskip}{0pt plus 3pt minus 2pt}
\end{figure}

RGBW is a new type of CFA (color filter array) pattern (Fig.~\ref{fig:rgbw_cfa}) designed for image quality enhancement under low light conditions. Thanks to the higher optical transmittance of white pixels over conventional red, green, and blue pixels, the signal-to-noise ratio (SNR) of images captured by this type of sensor increases significantly, thus boosting the image quality, especially under low light conditions. Recently, several phone OEMs~\cite{TranssionRGBW, oppoRGBW, vivoRGBW} have adopted RGBW sensors in their flagship smartphones to improve the camera image quality.

On the other hand, conventional camera ISPs can only work with Bayer patterns. Thereby, an interpolation procedure, which converts the CFA of the RGBW sensor into a Bayer pattern, is highly demanded. The interpolation procedure is usually referred to as remosaic (Fig.~\ref{fig:rgbw_cfa}), and a good remosaic algorithm should be able (1) to get a Bayer output from RGBW with the least artifacts and (2) to fully take advantage of the SNR and resolution benefit of white pixels.

The remosaic problem becomes more challenging when the input RGBW becomes noisy, especially under low-light conditions. A joint remosaic and denoise task is thus in demand for real-world applications.

\begin{figure*}[!ht]
\centering
\includegraphics[width=0.9\textwidth]{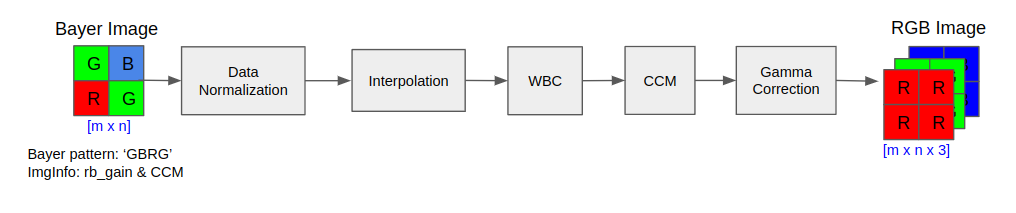}
\caption{An ISP to visualize the output Bayer and to calculate the loss function.}
\label{fig:simple_isp}
\setlength{\belowcaptionskip}{0pt plus 3pt minus 2pt}
\end{figure*}

In this challenge, we intend to remosaic the RGBW input to obtain a Bayer at the same spatial resolution. The solution is not necessarily learning-based. However, we provide a high-quality dataset of aligned RGBW and Bayer pairs to facilitate learning-based methods development, including 100 scenes (70 scenes for training, 15 for validation, and 15 for testing). The dataset is similar to the one provided in the first MIPI challenge, while we replaced some similar scenes with new ones. We also provide a simple ISP for participants to get the RGB image results from Bayer for quality assessment. Fig.~\ref{fig:simple_isp} shows the pipeline of the simple ISP. The participants are also allowed to use other public-domain datasets. The algorithm performance is evaluated and ranked using objective metrics: Peak Signal-to-Noise Ratio (PSNR), Structural Similarity Index (SSIM)~\cite{ssim}, Learned Perceptual Image Patch Similarity (LPIPS)~\cite{lpips}, and KL-divergence (KLD). 

We hold this challenge in conjunction with the second MIPI Challenge which will be held on CVPR 2023. Similar to the first MIPI challenges~\cite{feng2023mipi,sun2023mipi,yang2023mipi,yang2023mipi2,yang2023mipi3}, we are seeking algorithms that fully take advantage of the SNR and resolution benefit of white pixels to enhance the final Bayer image. MIPI 2023 consists of four competition tracks:

\begin{itemize}
    \item \textbf{RGB+ToF Depth Completion} uses sparse and noisy ToF depth measurements with RGB images to obtain a complete depth map.
    \item \textbf{RGBW Sensor Fusion} fuses Bayer data and a monochrome channel data into Bayer format to increase SNR and spatial resolution.
    \item \textbf{RGBW Sensor Remosaic} converts RGBW RAW data into Bayer format so that it can be processed by standard ISPs.
    \item \textbf{Nighttime Flare Removal} is to improve nighttime image
quality by removing lens flare effects.
\end{itemize}


\section{MIPI 2023 RGBW Sensor Remosaic}

To facilitate the development of high-quality RGBW Remosaic solutions, we provide the following resources for participants:
\begin{itemize}
    \item A high-quality dataset of aligned RGBW and Bayer. We enriched the scenes compared to the first MIPI challenge dataset. As far as we know, this is the only dataset consisting of aligned RGBW and Bayer pairs; 
    \item A script that reads the provided raw data to help participants get familiar with the dataset;
    \item A simple ISP including basic ISP blocks to visualize the algorithm outputs and to evaluate image quality on RGB results;
    \item A set of objective image quality metrics to measure the performance of a developed solution.
\end{itemize}

\subsection{Problem Definition}
RGBW remosaic aims to interpolate the input RGBW CFA pattern to obtain a Bayer of the same resolution. The remosaic task is needed mainly because current camera ISPs usually cannot process CFAs other than the Bayer pattern. In addition, the remosaic task becomes more challenging when the noise level gets higher, thus requiring more advanced algorithms to avoid image quality artifacts. Besides, RGBW sensors are widely used in smartphones with limited computational budgets and battery life, thus requiring the remosaic algorithm to be lightweight at the same time. While we do not rank solutions based on running time or memory footprint, the computational cost is one of the most important criteria in real applications.

\subsection{Dataset: Tetras-RGBW-Remosaic}
\begin{figure*}[!ht]
\centering
\includegraphics[width=0.9\textwidth]{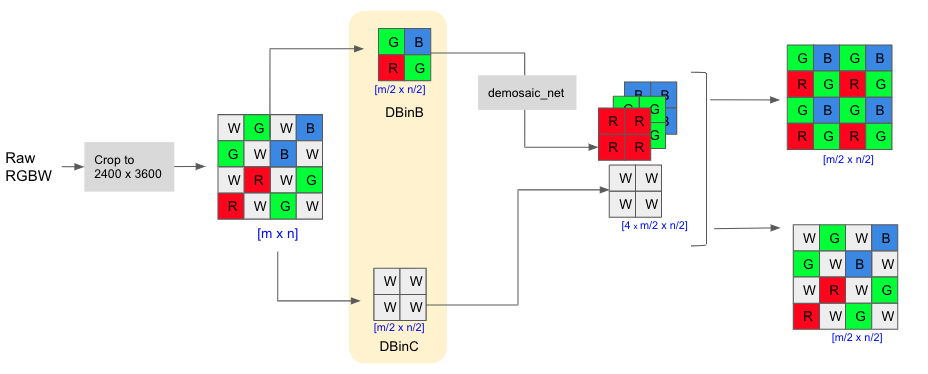}
\caption{Data generation of the RGBW remosaic task. The RGBW raw data is captured using a RGBW sensor and cropped into a size of $2400\times3600$. A Bayer (DBinB) and white (DBinC) image are obtained by averaging the same color in the diagonal direction within a $2\times2$ block. We demosaic the Bayer (DBinB) to get an RGB using the DemosaicNet~\cite{gharbi2016deep}. The white (DBinC) is concatenated to the RGB image to have RGBW for each pixel, which is in turn mosaiced to get the input RGBW and aligned ground truth Bayer.}
\label{fig:data_gen}
\setlength{\belowcaptionskip}{0pt plus 3pt minus 2pt}
\end{figure*}

The training data contains 70 scenes of aligned RGBW (input) and Bayer (ground truth) pairs. For each scene, noise is synthesized on the 0dB RGBW input to provide the noisy RGBW input at 24dB and 42dB, respectively. The synthesized noise consists of read noise and shot noise, and the noise models are calibrated on an RGBW sensor. The data generation steps are shown in Fig.~\ref{fig:data_gen}. The testing data includes RGBW inputs of 15 scenes at 0dB, 24dB, and 42dB, and the ground-truth Bayer results are hidden from participants during the testing phase.

\subsection{Evaluation}
The evaluation consists of (1) the comparison of the remosaic output Bayer and the reference ground truth Bayer, and (2) the comparison of RGB from the predicted and ground truth Bayer using a simple ISP (the code of the simple ISP is provided). We use
\begin{enumerate}
    \item Peak Signal-to-Noise Ratio (PSNR)
    \item Structural Similarity Index Measure (SSIM)~\cite{ssim}
    \item Learned Perceptual Image Patch Similarity (LPIPS)~\cite{lpips}
    \item Kullback–Leibler Divergence (KLD)
\end{enumerate}
to evaluate the remosiac performance. The PSNR, SSIM, and LPIPS will be applied to the RGB from the Bayer using the provided simple ISP code, while KLD is evaluated on the predicted Bayer directly.

A metric weighting PSNR, SSIM, KLD, and LPIPS is used to give the final ranking of each method, and we will report each metric separately as well. The code to calculate the metrics is provided. The weighted metric is shown below. The M4 score is between 0 and 100, and the higher score indicates the better overall image quality.

\begin{equation}
    M4 = PSNR \cdot SSIM \cdot 2^{1-LPIPS-KLD}.
\label{eq:M4}
\end{equation}
For each dataset, we report the average score over all the processed images belonging to it.

\subsection{Challenge Phase}
The challenge consisted of the following phases:
\begin{enumerate}
    \item Development: The registered participants get access to the data and baseline code, and are able to train the models and evaluate their running time locally.
    \item Validation: The participants can upload their models to the remote server to check the fidelity scores on the validation dataset, and to compare their results on the validation leaderboard.
    \item Testing: The participants submit their final results, code, models, and factsheets.
\end{enumerate}


\section{Challenge Results}

Table~\ref{tab:results} shows the top three teams' results. In the final test phase, we verified their submission using their code. \textbf{RUSH MI}, \textbf{HSTT}, and \textbf{MegNR} are the top three teams ranked by M4 as presented in Eq.~\eqref{eq:M4}, and \textbf{RUSH MI} shows the best overall performance. The proposed methods are described in Section \ref{sec:methods}, and the team members and affiliations are listed in Appendix \ref{appendix:teams}.

\begin{table}[!ht]
    \centering
    \scalebox{0.9}{
        \begin{tabular}{l | llll | l}
        \hline
            \textbf{Team name} & \textbf{PSNR} & \textbf{SSIM} & \textbf{LPIPS} & \textbf{KLD} & \textbf{M4}   \\ 
            \hline  \hline
            \text{RUSH MI}     & 38.545        & 0.976         & 0.0707         & 0.0650       & 68.72         \\
            \hline
            \text{HSTT}        & 38.739        & 0.974         & 0.0810         & 0.0669       & 68.51         \\
            \hline
            \text{MegNR}       & 38.004        & 0.965         & 0.0671         & 0.0684       & 67.10         \\
            \hline
        \end{tabular}
    }
    \caption{MIPI 2023 Joint RGBW Remosaic and Denoise challenge results and final rankings. PSNR, SSIM, LPIPS, and KLD are calculated between the submitted results from each team and the ground truth data. A weighted metric, M4, by Eq.~\eqref{eq:M4} is used to rank the algorithm performance, and the top three teams with the highest M4 are included in the table.  
    \label{tab:results}}
\end{table}

\begin{figure*}[!ht]
\setlength{\abovecaptionskip}{0.cm}
\setlength{\belowcaptionskip}{-0.cm}
\centering
\includegraphics[width=\textwidth]{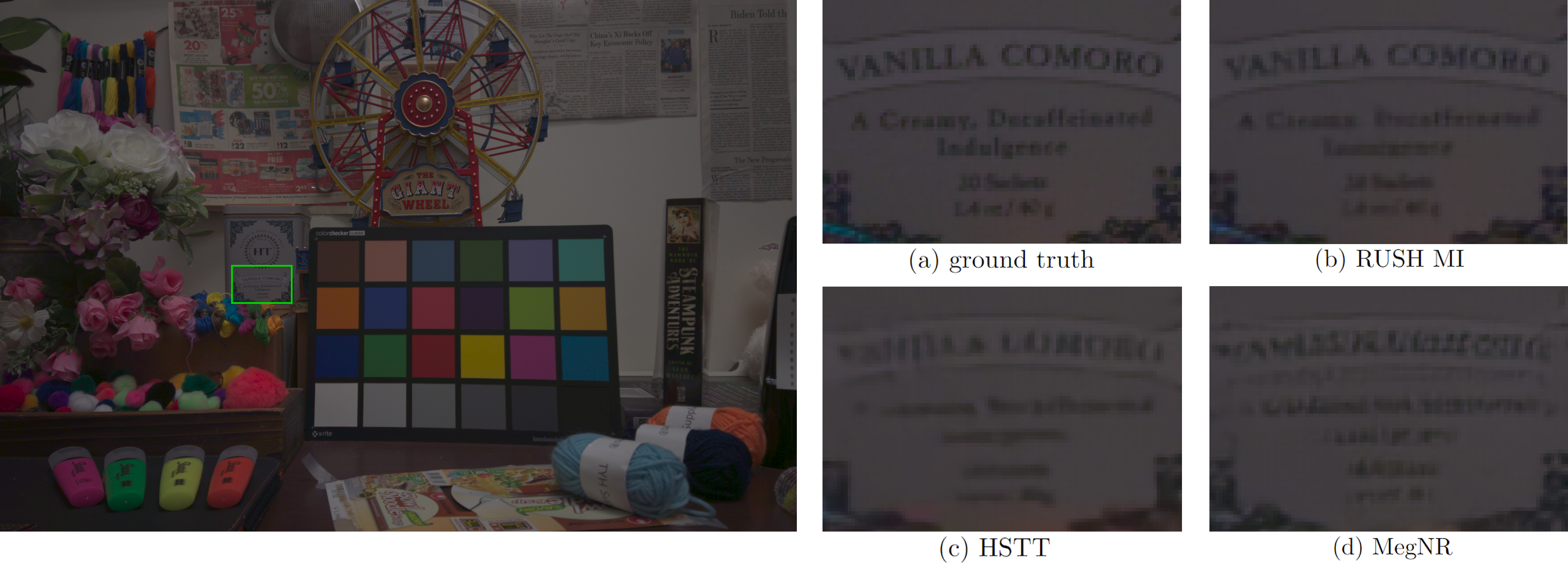}
\caption{Qualitative image quality (IQ) comparison. The results of one of the test scenes (42dB) are shown. While the top three remosaic methods achieve high objective IQ metrics in Table~\ref{tab:results}, texts on the metal box are slightly blurred in (b) and are barely interpretable in (c) and (d). The RGB images are obtained by using the ISP in Fig.~\ref{fig:simple_isp}, and its code is provided to participants.}
\label{fig:IQ1}
\end{figure*}

\begin{figure*}[!ht]
\setlength{\abovecaptionskip}{0.cm}
\setlength{\belowcaptionskip}{-0.cm}
\centering
\includegraphics[width=\textwidth]{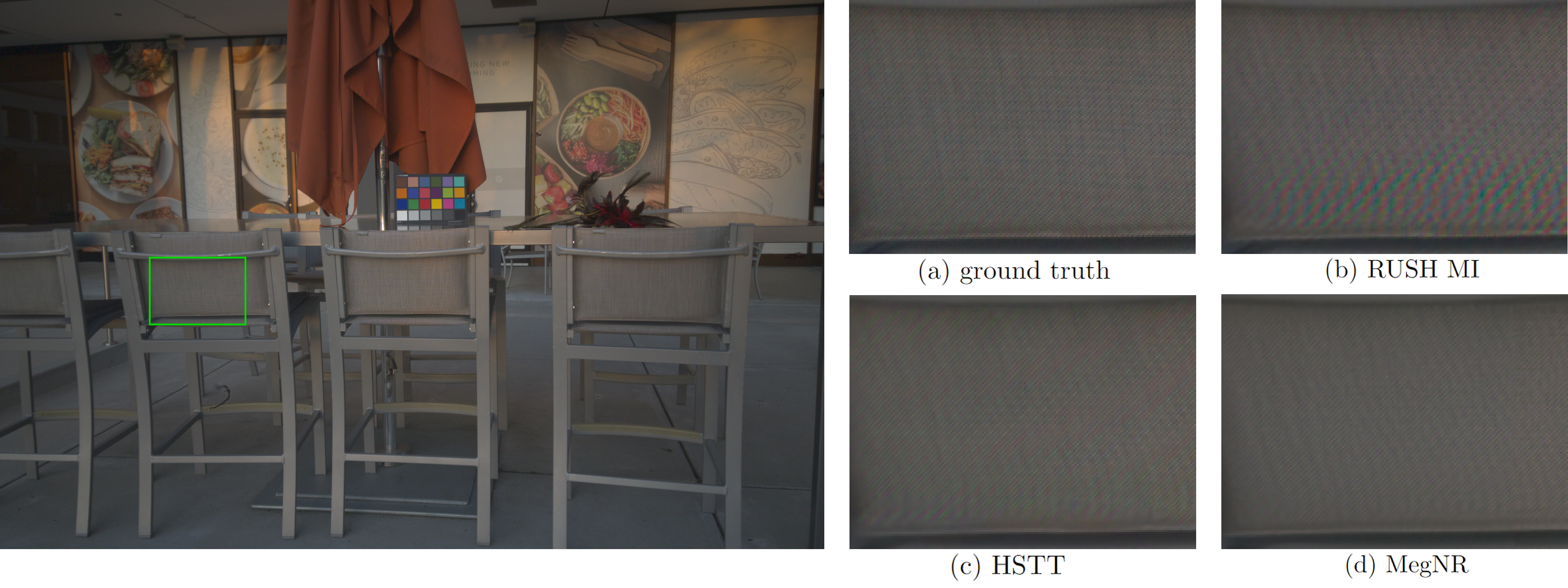}
\caption{Qualitative image quality (IQ) comparison. The results of one of the test scenes (42dB) are shown. Detail loss or false color in the top three methods in Table~\ref{tab:results} can be found when compared with the ground truth (a). There is slight detail loss on the mesh of the chair in (c) and (d) and some false color can be found in (b). The RGB images are obtained by using the ISP in Fig.~\ref{fig:simple_isp}, and its code is provided to participants.}
\label{fig:IQ2}
\end{figure*}

To learn more about the algorithm performance, we evaluated the qualitative image quality in Fig.~\ref{fig:IQ1} and Fig.~\ref{fig:IQ2} in addition to the objective IQ metrics. While all teams in Table~\ref{tab:results} have achieved high PSNR and SSIM, detail loss can be found on the texts of the metal box in Fig.~\ref{fig:IQ1} and detail loss or false color can be found on the mesh of the chair in Fig.~\ref{fig:IQ1}. When the input has a large amount of noise, oversmoothing tends to yield higher PSNR at the cost of detail loss perceptually.  

\begin{table}[!ht]  
    \centering
    \scalebox{0.9}{
        \begin{tabular}{l | l | l}
        \hline
            \textbf{Team name} & \textbf{1200$\times$1800 (measured)}  &  \textbf{64M} (estimated)\\ 
            \hline  \hline
            \text{RUSH MI}     & \textbf{0.26s}                        &  \textbf{7.7s}          \\ 
            \hline
            \text{HSTT}        & 73.31s                                &  2172s          \\ 
            \hline
            \text{MegNR}       & 6.02s                                 &  178s           \\ 
            \hline
    
        \end{tabular}
    }
    \caption{Running time of the top three solutions ranked by Eq.~\eqref{eq:M4} in the 2023 Joint RGBW Remosaic and Denoise challenge. The running time of input of $1200\times1800$ was measured, while the running time of a 64M input RGBW was based on the estimation. The measurement was taken on an NVIDIA Tesla V100-SXM2-32GB GPU.
    \label{tab:runtime}}
\end{table}

In addition to benchmarking the image quality of remosaic algorithms, computational efficiency is evaluated because of wide adoptions of RGBW sensors on smartphones. We measured the runnnig time of the remosaic solutions of the top three teams (based on M4 by Eq.~\eqref{eq:M4}) in Table~\ref{tab:runtime}. While running time is not employed in the challenge to rank remosaic algorithms, the computational cost is critical when developing smartphone algorithms. RUSH MI achieved the shortest running time among the top three solutions on a workstation GPU (NVIDIA Tesla V100-SXM2-32GB). With sensor resolution of mainstream smartphones reaching 64M or even higher, power-efficient remosaic algorithms are highly desirable.


\section{Challenge Methods}\label{sec:methods}

This section describes the solutions submitted by all teams participating in the final stage of the MIPI 2023 RGBW Joint Remosaic and Denoise Challenge. 

\subsection{RUSH MI}

\begin{figure}[!ht]
    \centering
     \includegraphics[width=\linewidth]{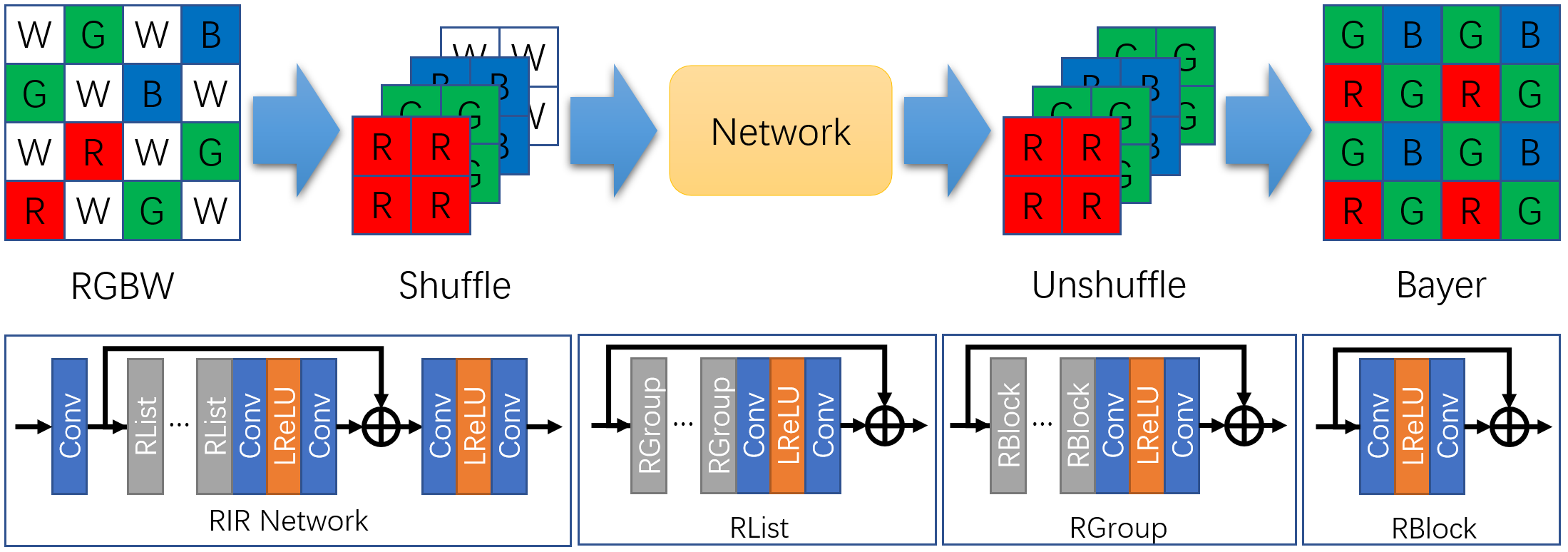}
    \caption{The network structure of Team RUSH MI.}
    \label{fig:rushmi}
\end{figure}

This team designs a residual-in-residual (RIR) network for the RGBW Joint Denoising and Remosaicing task. Fig.~\ref{fig:rushmi} shows the proposed network structure. This team notices that the R/G/B/W channels of RAW data have different representations on the spatial space. Directly applying the convolution to the RAW data will wrongly consider the spatial relationship among different color channels. To address this issue, this work shuffles the RGBW data into 16 channels and uses a RIR network for joint remosaicing and denoising. After that, the output 16-channel data is unshuffled to the GBRG bayer pattern. The basic component of this network is the residual block (RBlock) with two convolutional layers and one ReLU activation. The residual group (RGroup) is composed of several RBlocks, two convolutional layers and one ReLU activation. The residual list (RList) is composed of several RGroups, two convolutional layers and one ReLU activation. Finally, the backbone of RIR network is composed of several RLists, two convolutional layers and one ReLU activation. There is one convoluitonal layer exploring features from the shuffled RAW data. After restoration, the processed features are restored to RAW data by two convolutional layers and one ReLU activation. In the training phase, the clean model is used as a guidance for boosting the noisy restoration performance. The network is updated by Adam optimizer with initial learning rate as $10^{-4}$, which is halved for every 5000 iterations. The loss functions is chosen as L1 loss between the restored noisy RAW data and the clean label. The input data is randomly flipped and rotated for augmentation. 

\subsection{HSTT}

\begin{figure*}[!ht]
\centering
\begin{minipage}{0.95\linewidth}
    \centering
    \begin{minipage}{0.66\linewidth}
    \centering
        \includegraphics[width=\linewidth]{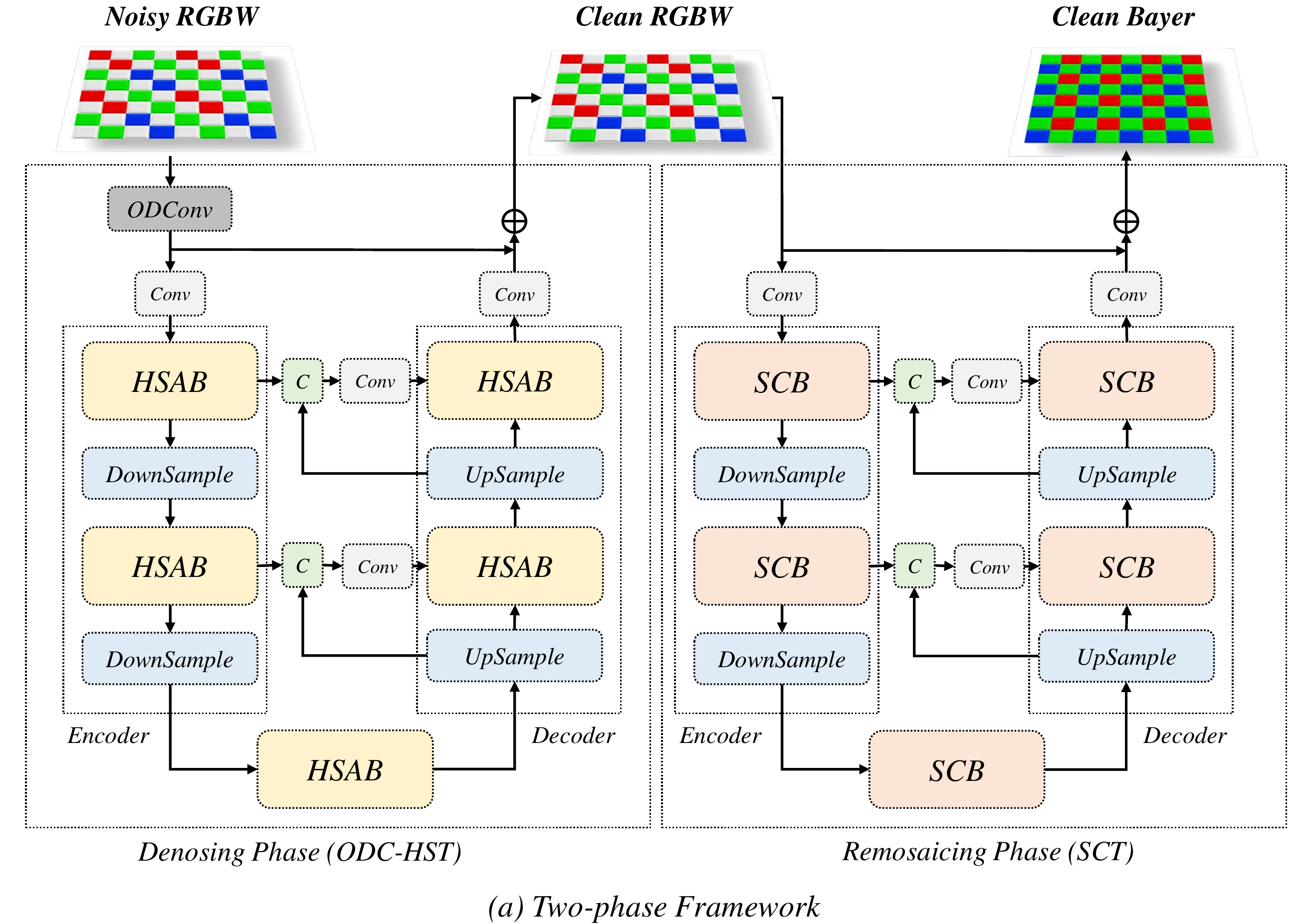}
    \end{minipage}
    \hspace{1mm}
    \begin{minipage}{0.3\linewidth}
    \centering
        \includegraphics[width=\linewidth]{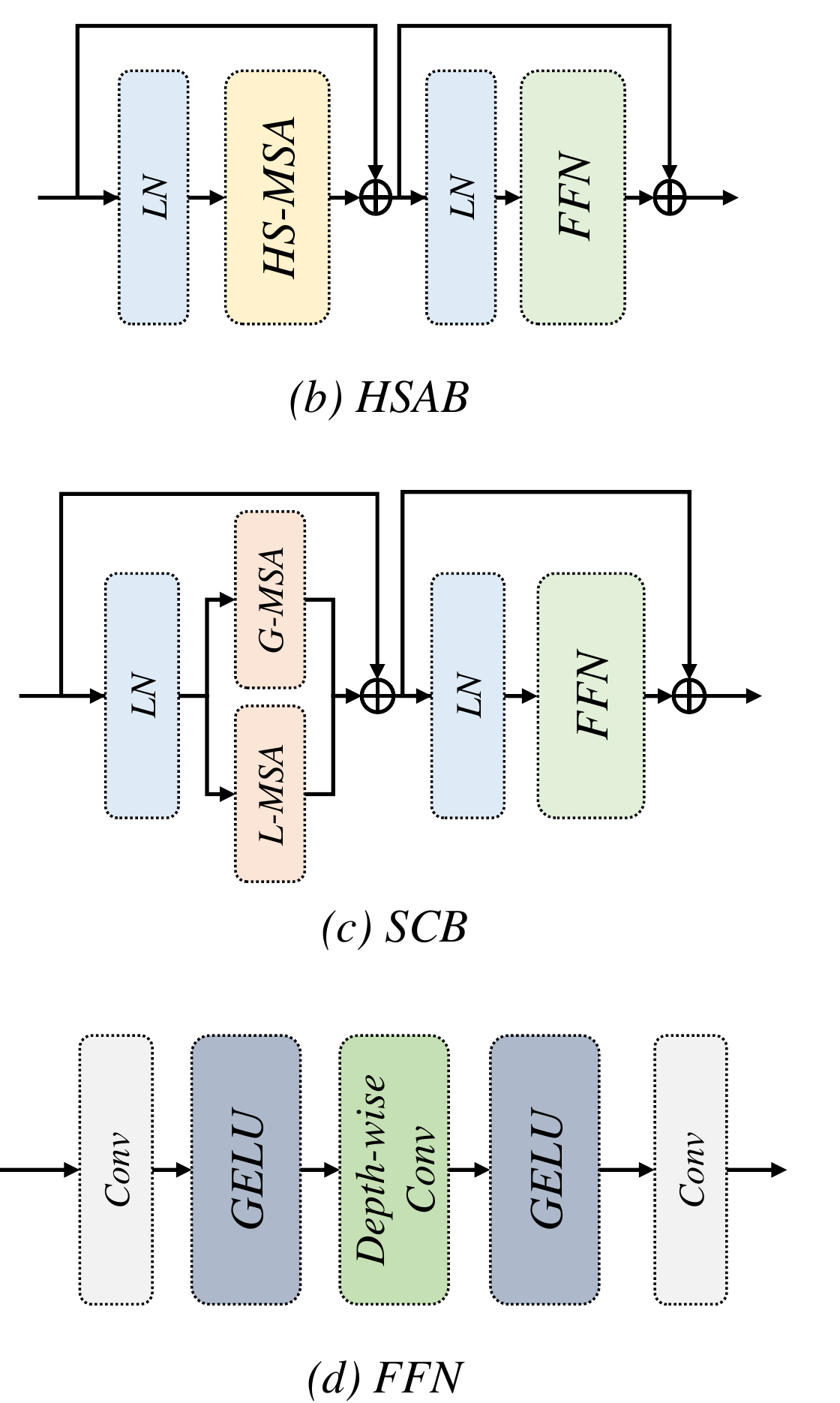}
    \end{minipage}
\end{minipage}
\caption{(a)~Illustration of our proposed two-phase RGBW-JDR framework OTST. The full framework consists of two sequential phases,~\ie, the denoising phase and remosaicing phase. Each phase contains a U-shaped structure Transformer. (b)~HSAB consists of an FFN, an HS-MSA, and two layer normalization. (c)~SCB consists of an FFN,  two layer normalization, parallel-connected L-MSA and G-MSA. (d)~Components of FFN.}
\label{fig_HSTT:framework}
\end{figure*}

\begin{figure*}[!ht]
\centering
\includegraphics[width=\linewidth]{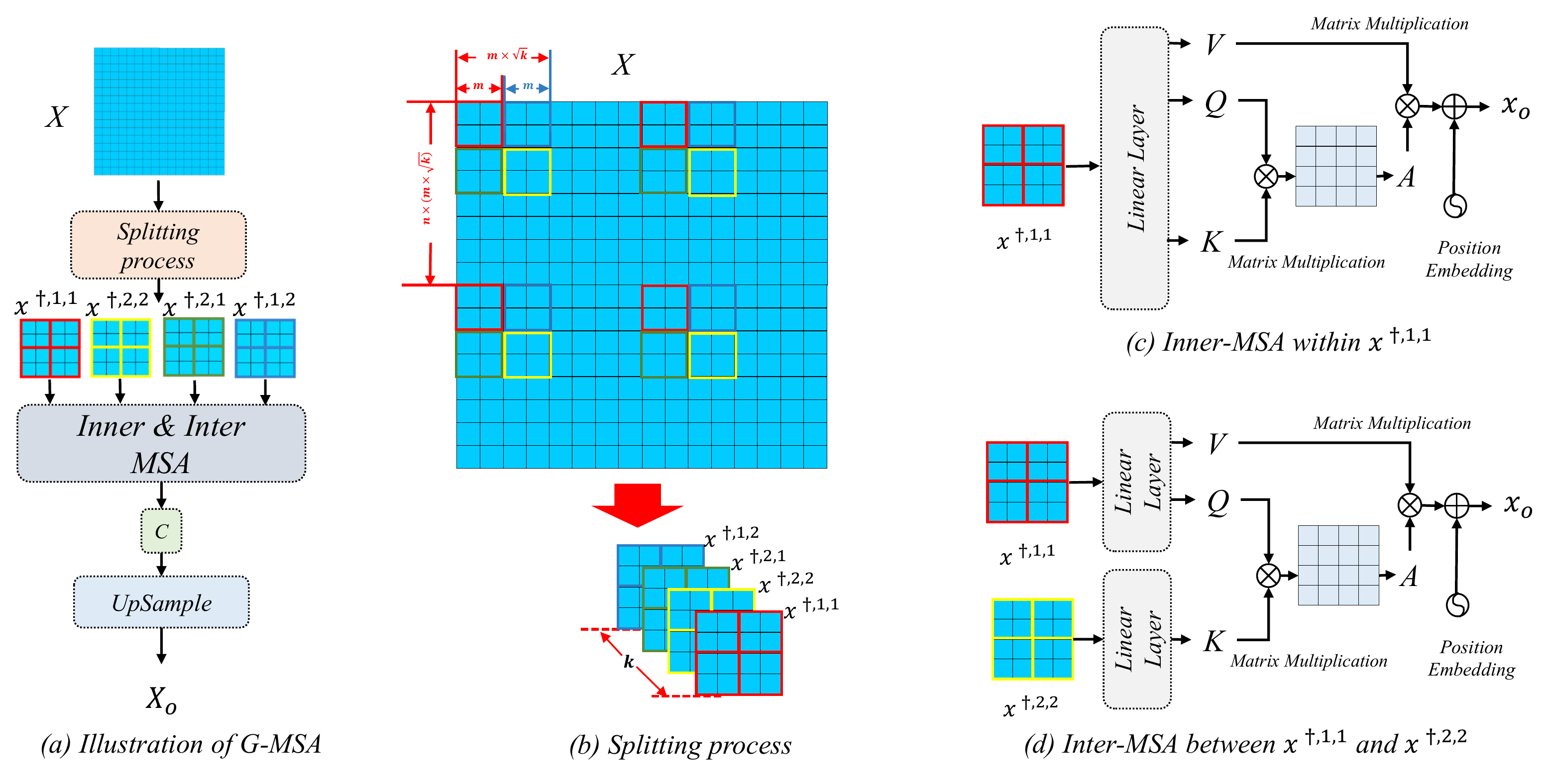}
\caption{(a) Visual illustration of G-MSA (b) Splitting process: G-MSA first samples $k$ visual tokens from input $\boldsymbol{X}$ with a dilation rate of $n=2$. Tokens with the same color borders belong to the same partition $\boldsymbol{x}^{\dag, i, j}$. (c) Illustration of inner-MSA for $\boldsymbol{x}^{\dag, 1, 1}$. (d) Illustration of inter-MSA between $\boldsymbol{x}^{\dag, 1, 1}$ and $\boldsymbol{x}^{\dag, 2, 2}$.} 
\label{fig_HSTT:G-MSA-L-MSA}
\end{figure*}

This team designs a two-phase framework named OTST for the RGBW Joint Denoising and Remosaicing (RGBW-JRD) task. For the denoising stage, we propose Omni-dimensional Dynamic Convolution based Half-Shuffle Transformer (ODC-HST) which can fully utilize image's long-range dependencies to dynamically remove the noise. For the remosaicing stage, we propose a Spatial Compressive Transformer (SCT) to efficiently capture both local and global dependencies across spatial and channel dimensions. The entire framework is shown in the Fig.~\ref{fig_HSTT:framework}.
The whole two-phase framework can be formulated as:
\begin{equation}
\boldsymbol{X} = \mathbf{F}_{\odot} \bigg( \mathbf{F}_{\gamma} \Big(\boldsymbol{Y} + \mathcal{N}\left(0,\ \boldsymbol{Y} \cdot \sigma^2_{s} + \sigma^2_{c}\right) | \boldsymbol{\theta}_{\gamma} \Big) | \boldsymbol{\theta}_{\odot} \bigg).
\end{equation}
Where 
$\boldsymbol{\theta}_{\odot},~\boldsymbol{\theta}_{\gamma}$ denote the learnable parameters in $\mathbf{F}_{\odot}$ and $\mathbf{F}_{\gamma}$. $\boldsymbol{Y}$ represents the input and $\mathcal{N}$ represents the noise distribution. $\boldsymbol{X}$ represents the final output image of Bayer.

Specifically, for the denoising phase, This team designs ODC-HST to play the role of denoising, which consists of two sequential modules: an Omni-dimensional Dynamic Convolution~(ODC)~\cite{li2022omni} to obtain the noise distribution of the entire raw image, and a Half-Shffle Transformer~(HST)~\cite{cai2022degradation} to eliminates the noise.

For the remosaicing phase, This team designs the Spatial Compressive Transformer~(SCT), which aims to efficiently model both local-global spatial self-similarities and inter-channel correlation. As shown in Fig.~\ref{fig_HSTT:framework}~(c), to achieve this, the basic unit of SCT, named Spatial Compressive Block (SCB), a Local-Global Dual Spatial-wise MSA~(DS-MSA) module  to capture both local high-frequency details and long-range global dependencies at the same time. The overall processes of G-MSA is shown in Fig.~\ref{fig_HSTT:G-MSA-L-MSA}~(a). Generally speaking, G-MSA first splits input features into several dilated partitions along spatial dimension. As shown in Fig.~\ref{fig_HSTT:G-MSA-L-MSA}~(b), it is worth nothing that pixels in each partition are not from a local region but subsampled from the whole input feature with a dilation rate. Then G-MSA computes both inner-MSA and inter-MSA between each pairs of partitions to capture global dependencies. Fig.~\ref{fig_HSTT:G-MSA-L-MSA}~(c) $\sim$ (d) provide examples by computing inner-MSA and inter-MSA, respectively. After that, G-MSA concatenates these outputs along channel dimension. Finally, a upsample$\times 2$ module is employed to scale up the acquired features to match the spatial dimensionality of the original input.

By aggregating the outputs of the L-MSA and G-MSA branches, our DS-MSA achieves the ability to efficiently capture both local and global spatial information simultaneously.

MAE loss functions are used in denoise and remosaic:

\begin{equation}
\left\{\begin{matrix}
\mathcal{L}_{D}  = \Vert \boldsymbol{X}_{*} - \boldsymbol{X}_{0}\Vert_1 
 \\
\mathcal{L}_{R}  = \Vert \boldsymbol{X} - \boldsymbol{I}_{gt} \Vert_1 + \lambda \Vert \boldsymbol {X}_{rgb} - \boldsymbol I_{rgb} \Vert_1
\end{matrix}\right.
\end{equation}
Where $\boldsymbol{X}_{*}$ and $\boldsymbol{X}_{0}$ denotes the clean output of denose phase and 0dB RGBW. $\boldsymbol{X}$ and $\boldsymbol{I}_{gt}$ represent the reconstructed Bayer of remosaic model and ground truth Bayer respectively. $\boldsymbol{X}_{rgb}$ and $\boldsymbol{I}_{rgb}$ indicates $\boldsymbol{X}$ and $\boldsymbol{I}_{gt}$ after the official ISP to obtain RGB images. $\lambda$ is a hyper-parameter tuning $\mathcal{L}_{R}$.

The training details are presented as follows: the model is implemented in Pytorch and performed on 8 Titan XP graphical processing units (GPUs). The model is optimized using an Adam optimizer with parameters  $\beta_1 = 0.9$, $\beta_2 = 0.99$, learning rate = $1e-4$ with a batch size of $5$ and a patch size of $128$.

\subsection{MegNR}

\begin{figure*}[!ht]
    \centering
    \includegraphics[width=0.9\linewidth]{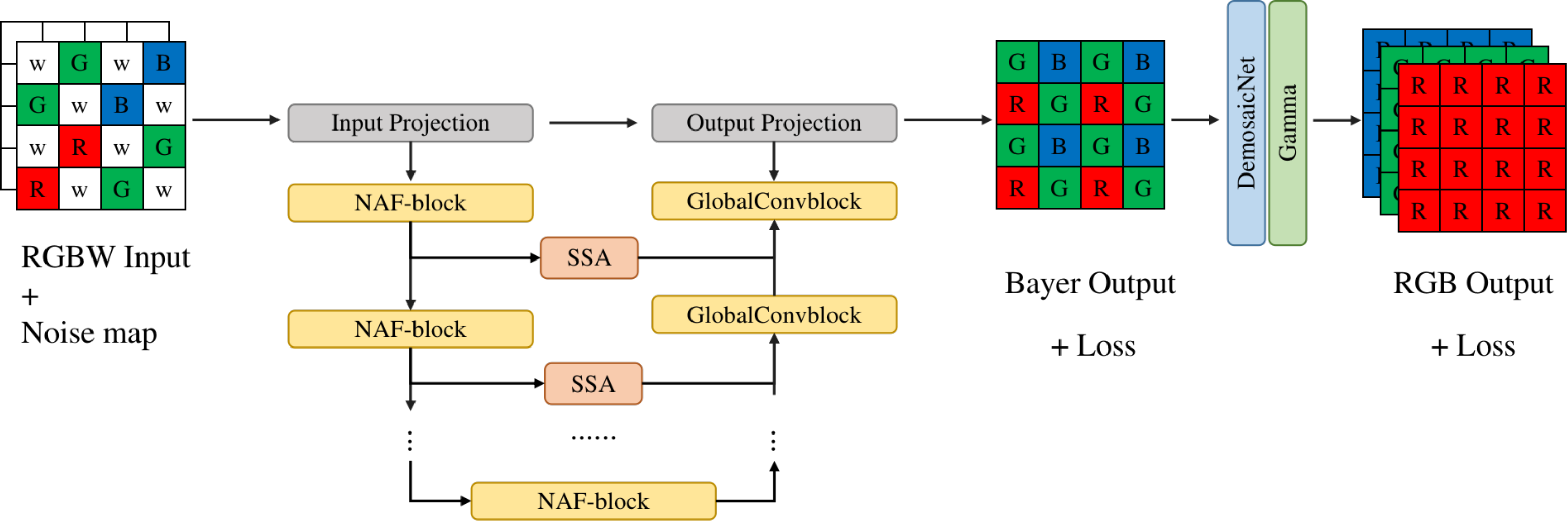}
    \caption{The overall pipeline of our method and the structure of SNN.}\label{fig:naf_ssa_pipeline}
\end{figure*}

The proposed method for RGBW remosaicking is based on NAFNet~\cite{chen2022simple} and the SSA module~\cite{cheng2021nbnet}, so-called SNN (Subspace Nonlinear Network). As show in Fig.\ref{fig:naf_ssa_pipeline}, the network architecture consists of a U-shaped encoder-decoder with 4 stages, where the encoder is composed of multiple NAF blocks~\cite{chen2022simple}  while the decoder has multiple global convolution blocks as depicted in the NBNet paper~\cite{cheng2021nbnet}. The encoder takes the concatenated RGBW image and noise map as input and extracts features using the NAF blocks. Specifically, the numbers of NAF blocks for each stage in the encoder are 2, 2, 4, and 8. The decoder takes the encoded information and reconstructs the GBRG Bayer image. It also consists of 4 stages, with 12 NAF blocks between the encoder and decoder, and 2 global convolution blocks for each stage.

To improve feature separation and noise removal, the SSA module is introduced to replace the original upsampling structure. The SSA module utilizes low-level features from the encoder and high-level features from the corresponding decoder stage to better separate information and noise. This module helps reduce residual noise and artifacts in the output.

After obtaining the recovered GBRG Bayer image, the official simple ISP code is used to convert it to an RGB image. The loss is then computed on both the Bayer and RGB results. In addition to the traditional MSE loss, a new loss function that incorporates visual perception metrics such as PSNR, SSIM, and LIPIPS is proposed. These metrics are commonly used in image processing tasks and provide a more meaningful evaluation of image quality compared to just using the MSE loss.

Overall, the approach focuses on improving feature separation and noise reduction in the RGBW remosaicking process to achieve better image quality. The use of NAFNet and the SSA module allows accomplishing this while maintaining an end-to-end framework. Additionally, the incorporation of visual perception metrics in the loss function enables evaluating the performance of the model in a more meaningful way.


\section{Conclusions}
This report reviewed and summarized the methods and results of RGBW Remosaic challenge in the 2nd Mobile Intelligent Photography and Imaging workshop (MIPI 2023) held in conjunction with CVPR 2023. 
The participants were provided with a high-quality dataset for RGBW Remosaic and denoising.
The top three submissions leverage learning-based methods and achieve promising results.
We are excited to see so many submissions within such a short period, and we look forward to more research in this area.


\section{Acknowledgements}
We thank Shanghai Artificial Intelligence Laboratory, Sony, and Nanyang Technological University for sponsoring this MIPI 2023 challenge. We thank all the organizers and participants for their great work.


{\small
\bibliographystyle{ieee_fullname}
\bibliography{egbib}
}

\appendix

\section{Teams and Affiliations}
\label{appendix:teams}

\subsection*{RUSH MI}
\noindent
\textbf{Title}:\\ 
Residual-in-Residual Network for Joint RGBW data Denoising and Remosaicing \\
\textbf{Members}:\\
Yuqing Liu$^1$ (\href{lyqatdl@163.com}{lyqatdl@163.com}),\\
Hongyuan Yu$^2$, Weichen Yu$^{3}$, Zhen Dong$^{2}$, Binnan Han$^{2}$, Qi Jia$^{1}$, Xuanwu Yin$^{2}$, Kunlong Zuo$^{2}$\\
\textbf{Affiliations}:\\
$^1$ Dalian University of Technology,\\
$^2$ Institute of Automation, Chinese Academy of Sciences,\\
$^3$ Xiaomi Inc.
\\

\subsection*{HSTT}
\noindent
\textbf{Title}:\\ 
OTST: A Two-Phase Framework for Joint Denoising and Remosaicing in RGBW CFA \\
\textbf{Members}:\\
Yaqi Wu$^1$ (\href{titimasta@163.com}{titimasta@163.com}),\\
Zhihao Fan$^{1,2}$, Fanqing Meng$^{3}$, Xun Wu$^{1,4}$, Jiawei Zhang$^1$, Feng Zhang$^1$\\
\textbf{Affiliations}:\\
$^1$ Tetras.AI,\\
$^2$ University of Shanghai for Science and Technology,\\
$^3$ Tongji University,\\
$^4$ Tsinghua University
\\

\subsection*{MegNR}
\noindent
\textbf{Title}:\\ 
SNN: Subspace Nonlinear Network for Joint RGBW data Denoising and Remosaicing \\
\textbf{Members}:\\
Mingyan Han$^1$ (\href{hanmingyan@megvii.com}{hanmingyan@megvii.com}),\\
Jinting Luo$^1$, Qi Wu$^1$, Ting Jiang$^1$, Chengzhi Jiang$^1$, Xinpeng Li$^1$,  Wenjie Lin$^1$, Lei Yu$^1$, Haoqiang Fan$^1$, Shuaicheng Liu$^{2, 1*}$\\
\textbf{Affiliations}:\\
$^1$ Megvii Technology,\\
$^2$ University of Electronic Science and Technology of
China (UESTC)\\

\end{document}